\documentclass{edbk} 
\usepackage{edbkps}

\normallatexbib

\begin{document}



\articletitle[ Multipartite 
Greenberger-Horne-Zeilinger paradoxes for continuous variables]{ Multipartite 
Greenberger-Horne-Zeilinger paradoxes for continuous variables}

\chaptitlerunninghead{Multipartite GHZ paradoxes}

\author{Serge Massar and Stefano Pironio\\
Service de Physique Th\'eorique, CP 225,\\
Universit\'e Libre de Bruxelles, 1050 Brussels, Belgium \\
}



\begin{abstract}
We show how to construct 
Greenberger-Horn-Zeilinger type paradoxes
for continuous variable systems.  We give two examples corresponding
to 3 party and 5 party paradoxes. The paradoxes are revealed by carrying
out position and momentum measurements. The structure of the quantum
states which lead to these paradoxes is discussed.
\end{abstract}



When studying continuous variables systems, described by conjugate 
variables with commutation relation $[x,p]=i$, it is natural to inquire how 
non-locality can be revealed in those systems. Experimentally the
operations that are easy to carry out on such systems involve linear optics,
squeezing and homodyne detection. Using these operations, the states that can
be prepared are Gaussian states and the measurements that can be performed
are measurements of quadratures. But Gaussian states possess a Wigner function
which is positive everywhere and so provide a trivial local-hidden variable
model for measurement of $x$ or $p$.

To exhibit non-locality in these systems, it is thus necessary to drop some of
the requirements imposed by current day experimental techniques. 
For instance one can invoke
more challenging measurements such as photon counting measurements or 
consider more general states that will necessitate higher order 
non-linear couplings to be
produced.  Using these two approaches it has recently been possible to
extend from discrete variables to continuous variables systems the usual
non-locality tests: Bell inequalities \cite{Bana, Kuz, Chen}, Hardy's
non-locality proof \cite{Hill} and the Greenberger-Horne-Zeilinger paradox
\cite{C,MP, ChenGHZ}.

Greenberger-Horne-Zeilinger (GHZ) paradoxes \cite{GHZ} as formulated by
Mermin \cite{M} are particularly elegant and simple ways of demonstrating the
non-locality of quantum systems since the argument can be carried out
at the level of operators only. The existence of a generalization of the
original GHZ paradox for qubits to continuous variables was first pointed out by
Clifton \cite{C} and was studied in more details in \cite{MP} and
\cite{ChenGHZ}. The paradox presented in \cite{ChenGHZ} 
involve measurements of the parity of the number of photons, while in \cite{C} 
and \cite{MP}, it is associated with position and momentum variables. It is 
this last case that we will consider here. We shall summarize the results of 
\cite{MP} and show that the multipartite multidimensional GHZ paradoxes 
introduced in \cite{CMP} can easily be generalized to the case of 
continuous variables by exploiting the noncommutative geometry of the 
phase space. This idea is closely related to the technique used to embed 
finite-dimensional quantum error correcting code in the infinite-dimensional 
Hilbert space of continuous variables systems \cite{Got}.

Let us introduce the dimensionless variables
\begin{equation}
\tilde x = {x \over {\sqrt{\pi}L}} \quad \mbox{and} \quad
\tilde p = {p\ L \over \sqrt{\pi}} \ ,
\end{equation}
where $L$ is an arbitrary length scale. Consider the translation operators in phase space 
\begin{equation}
X^\alpha = \exp ( i \alpha \tilde x) \quad \mbox{and} \quad
Y^\beta = \exp ( i \beta \tilde p) \ .
\label{XY}
\end{equation} 
These unitary operators obey the commutation relation
\begin{equation}\label{commut}
X^\alpha Y^\beta =e^{i\alpha \beta/\pi} Y^\beta X^\alpha \ ,
\end{equation}
which follows from $[\tilde x,\tilde p]=i/\pi$ and the identy 
$e^Ae^B=e^{[A,B]}e^Be^A$ (valid if $A$ and $B$ commute with their commutator).
The continuous variable GHZ paradoxes will be built out of these operators. 

Let us first consider the case of three spatially separated parties, A, B, C, 
each of which possess one part of an entangled system described by the 
canonical variables $x_A,p_A, x_B, p_B, x_C$ and  $p_C$. Consider the operators
$X_j^{\pm \pi}  $ and $Y_j^{\pm \pi}$ acting on the space
of party $j$ ($j=A,B,C$).  Since $\alpha\beta=\pm \pi^2$, it follows from 
(\ref{commut}), that these operators obey the commutations relations 
$X_j^{\pm \pi} Y_j^{\pm \pi}=-Y_j^{\pm \pi}X_j^{\pm \pi}$.
Using these operators let us construct the following
four GHZ operators:
\begin{eqnarray}
\begin{array}{cccccc}
V_1 &=& X_A^\pi & X_B^\pi & X_C^\pi \\  
V_2 &=& X_A^{-\pi}&  Y_B^{-\pi} &Y_C^{\pi} \\
V_3 &=& Y_A^{\pi} &  X_B^{-\pi}&   Y_C^{-\pi} \\
V_4 &=& Y_A^{-\pi} & Y_B^{\pi} & X_C^{-\pi} \\
\label{V4}
\end{array}\end{eqnarray}
These four operators give rise to a GHZ paradox as we now show. First
note that 
the following two properties hold:
\begin{enumerate}
\item $V_1, V_2, V_3, V_4$ all commute. Thus they can be
  simultaneously diagonalized (in fact there exists a complete set
  of common eigenvectors).
\item The product $V_1 V_2 V_3 V_4 =
-I_{ABC} $ equals minus the identity operator.
 \end{enumerate}
These properties are easily proven using the commutations relations
$X_j^{\pm \pi} Y_j^{\pm \pi}=-Y_j^{\pm \pi}X_j^{\pm \pi}$. Any common 
eigenstate of $V_1, V_2, V_3, V_4$ will give rise to a GHZ paradox. Indeed 
suppose that the  parties measure the hermitian operators $x_j$ or
$p_j$, $j=A,B,C$ on this common eigenstate. The result of the
measurement associates a complex number of unit norm to either the
$X_j$ or $Y_j$ unitary operators.
If one of the  combinations of operators that occurs in eq. (\ref{V4}) is
measured, a value can be assigned to one of the operators $V_1, V_2, V_3, V_4$. 
Quantum mechanics imposes that this value is equal to the corresponding 
eigenvalue. Moreover - due to property 2 - the product of the
eigenvalues is $-1$. 

But this is in contradiction with local hidden variables theories. Indeed in
a local hidden theory one must assign, prior to the measurement, a complex
number of unit norm to all the operators $X_j$ and $Y_j$. Then taking the
product of the four c-numbers assigned simultaneously to $V_1, V_2,V_3, V_4$ 
yields $+1$ instead of $-1$.

Remark that all other tests of non-locality for continuous variable
systems
\cite{Bana,Kuz,Chen,Hill,ChenGHZ}
use measurements with a discrete spectrum (such as the parity photon
number) or  
involving only a discrete set of outcome (such as the probability that
$x>0$  
or $x<0$). In our version of the GHZ paradox for continuous variables this 
discrete character doesn't seem to appear at first sight. However it turn out 
that is is also the case thought in a subtle way because 
eq. (\ref{V4}) can be viewed as an infinite set of 2 dimensional
paradoxes (see \cite{MP} for more details).

In \cite{CMP}, GHZ paradoxes for many parties and multi-dimensional 
systems where constructed.  These paradoxes where build using
$d$-dimensional unitary operators with commutation relations:
\begin{equation}\label{discop}
X Y = e^{2\pi i/d} Y X
\end{equation}
which is a generalization of the anticommutation relation of spin operators
for two-dimensional systems. Using $X^{\alpha}$ and $Y^{\beta}$ and
choosing the  
coefficients $\alpha$ and $\beta$ such that $\alpha\beta=2\pi^2/d$
with $d$ integer, this 
commutation relation can be realized in a continuous variable systems
and so all  
the paradoxes presented in \cite{CMP} can be rephrased with minor
modifications in the context of 
infinite-dimensional Hilbert space.

Let us for instance generalise to continuous variables the paradox for 5
parties each having a 4 dimensional systems described in \cite{CMP}.
We now consider the operators $X^{\pm q}$, $Y^q$ and $Y^{-3q}$ where 
$q=\pi/\sqrt{2}$. They obey the commutation relation 
$X^{\pm q}Y^{q}= e^{\pm i \pi/2}Y^{q}X^{\pm q}$ and $X^{\pm q}Y^{-3q}= e^{\pm 
i \pi/2}Y^{-3q}X^{\pm q}$. Consider now the six unitary operators 
\begin{eqnarray}\begin{array}{ccccccc} 
W_1 &=& X^{q}_A &X^{q}_B& X^{q}_C & 
X^{q}_D &  X^{q}_E \\ 
W_2 &=& X^{-q}_A &Y^{-3q}_B & Y^{q}_C & Y^{q}_D&
Y^{q}_E \\
W_3 &=&  Y^{q}_A& X^{-q}_B&  Y^{-3q}_C & Y^{q}_D &
Y^{q}_E \\
W_4 &=& Y^{q}_A & Y^{q}_B& X^{-q}_C & Y^{-3q}_D&
Y^{q}_E \\
W_5 &=&  Y^{q}_A & Y^{q}_B & Y^{q}_C & X^{-q}_D &
Y^{-3q}_E \\
W_6 &=& Y^{-3q}_A & Y^{q}_B &Y^{q} _C&
Y^{q}_D& X^{-q}_E
\end{array}\label{VV6}
\end{eqnarray}
One easily shows that these six unitary operators commute and that their product
is minus the identity operator. Furthermore if one assigns a classical value to
$x_j$ and to $p_j$ for $j=A,B,C,D,E$, then the product of the operators takes
the value $+1$. Hence, using the same argument as in the three party case, we 
have a contradiction. 

There is a slight difference between the paradox (\ref{VV6}) 
and the 4-dimensional paradox described in \cite{CMP}. The origin of this 
difference is that in a $d$-dimensional Hilbert space, if unitary operators 
$X,Y$ obey $XY=e^{i\pi/d}YX$, then $X^d = Y^d = I$ (up to a phase which we set 
to 1). In the 4-dimensional case, this implies that $X^3 = X^\dagger$ and 
$Y^3=Y^\dagger$. In the continous case these relations no longer hold and the 
GHZ operators $W_i$'s must be slightly modified, i.e. the operator 
$X^{-q}={X^q}^{\dagger}$ and $Y^{-3q}={Y^{3q}}^{\dagger}$ have to be explicitly 
introduced in order for the product of the $W_i$'s to give minus the idendity.   
Note that the same remark apply for the previous paradox (\ref{V4}) where in the 
discrete 2-dimensional version $X^\dagger=X$ and $Y^\dagger=Y$.

As we mentioned earlier the GHZ state are not Gaussian states.
A detailed analysis of the common eigenstates of $V_1, V_2, V_3, V_4$
is given in \cite{MP}. Let us give an example of such an eigenstate.
Define the following coherent superpositions of infinitely squeezed states:
\begin{eqnarray}
|{\uparrow} \rangle
&=& {1 \over \sqrt{2}}
\sum_{k=-\infty}^{\infty}  \left (|{\tilde x =
2k}\rangle
+i
|{\tilde x= \tilde 2k+1}\rangle \right)
\nonumber\\
|{\downarrow} \rangle &=& {1 \over \sqrt{2}}
\sum_{k=-\infty}^{\infty}
\left(  |{\tilde x =2k} \rangle
-i |{\tilde x= \tilde
2k+1}\rangle \right)   \ ,
\end{eqnarray}
where $|\tilde x \rangle=|x=\sqrt{\pi}L\tilde x\rangle$.
Then a common eigenstate of the four GHZ operators $V_1,V_2,V_3,V_4$ is
$$\left (|\uparrow\rangle_A|\uparrow\rangle_B|\uparrow\rangle_C
-
|\downarrow\rangle_A|\downarrow\rangle_B|\downarrow\rangle_C\right)/\sqrt{2}\
.$$
However as shown in \cite{MP}, for any choice of the eigenvalues of
the operators $V_k$, there is an infinite family of eigenvectors,
ie. the eigenspace is infinitely degenerate.

In summary we have shown the existence of multidimensional and
multipartite GHZ paradoxes for continuous variable systems. These
paradoxes involve measurements of position and momentum variables
only, but the states which are measured are complex and difficult to
construct experimentaly.



\begin{acknowledgments}
We would like to thank N. Cerf for helpful discussions.
We acknowledges funding by the European
Union under project EQUIP (IST-FET program).
S.M. is a research associate of the Belgian
National Research Foundation. 
\end{acknowledgments}



%


\bibliographystyle{apalike}


\end{document}